# Fusion of Millimeter-wave Radar and Pulse Oximeter Data for Low-burden Diagnosis of Obstructive Sleep Apnea-Hypopnea Syndrome

Wei Wang, Zhaoxi Chen, Wenyu Zhang, Zetao Wang, Xiang Zhao, Chenyang Li, Jian Guan, Shankai Yin, Gang Li*, *Senior Member, IEEE*

*Abstract*—*Objective*: The aim of the study is to develop a novel method for improved diagnosis of obstructive sleep apnea-hypopnea syndrome (OSAHS) in clinical or home settings, with the focus on achieving diagnostic performance comparable to the gold-standard polysomnography (PSG) with significantly reduced monitoring burden. *Methods*: We propose a method using millimeter-wave radar and pulse oximeter for OSAHS diagnosis (ROSA). It contains a sleep apnea-hypopnea events (SAE) detection network, which directly predicts the temporal localization of SAE, and a sleep staging network, which predicts the sleep stages throughout the night, based on radar signals. It also fuses oxygen saturation ($SpO_2$) information from the pulse oximeter to adjust the score of SAE detected by radar. *Results*: Experimental results on a real-world dataset (>800 hours of overnight recordings, 100 subjects) demonstrated high agreement (ICC=0.9870) on apnea-hypopnea index (AHI) between ROSA and PSG. ROSA also exhibited excellent diagnostic performance, exceeding 90% in accuracy across AHI diagnostic thresholds of 5, 15 and 30 events/h. *Conclusion*: ROSA improves diagnostic accuracy by fusing millimeter-wave radar and pulse oximeter data. It provides a reliable and low-burden solution for OSAHS diagnosis. *Significance*: ROSA addresses the limitations of high complexity and monitoring burden associated with traditional PSG. The high accuracy and low burden of ROSA show its potential to improve the accessibility of OSAHS diagnosis among population.

*Index Terms* — fusion, millimeter-wave radar, pulse oximeter, OSAHS, deep learning

## I. Introduction

The obstructive sleep apnea-hypopnea syndrome (OSAHS) is among the most prevalent sleep-related breathing disorders worldwide [1]. It is characterized by repetitive upper airway obstruction and reduced airflow, resulting in compromised sleep quality [2]. Based on the underlying causes, sleep apnea-hypopnea events (SAE) are generally categorized into central, obstructive, mixed apnea (CA, OA, MA) and hypopnea (HP). Severe OSAHS can lead to a variety of comorbidities, including anxiety, stroke and hypertension, significantly affecting the patients' quality of life. According to review studies, global prevalence estimation for OSAHS in adults ranges from 9% to 38%, with rates exceeding 78% among the elderly [3].

The polysomnography (PSG) is a diagnostic tool of many types of sleep disorders, which is widely used in sleep medicine. It monitors multiple physiological signals of subjects during sleep [4], including respiratory airflow, respiratory efforts, pulse oxygen saturation ($SpO_2$), electromyogram (EMG), electrocardiogram (ECG), electrooculogram (EOG) and electroencephalogram (EEG), etc. Through analysis of these physiological signals by sleep technologists, PSG can provide comprehensive assessment of sleep status, and then diagnose sleep disorders such as OSAHS, insomnia, etc. However, PSG requires patients to be monitored in a clinical setting, where they are fitted with multiple contact sensors throughout the night. These sensors often bring significant discomfort, leading to poor sleep quality. Moreover, the first-night effect [5] may result in altered sleep patterns, leading to deviations from normal sleep and affecting diagnostic accuracy. These factors limit the widespread adoption of PSG, reducing the accessibility of accurate diagnosis for broader population.

The apnea-hypopnea index (AHI) is a key indicator to diagnose OSAHS in clinical practice [6]. It is defined as the total number of apnea and hypopnea events occurring per hour of sleep. These breathing events are identified by analyzing respiratory airflow, respiratory effort and $SpO_2$, while total sleep time (TST) is mainly derived from the analysis of EEG, EOG and EMG. Radar, known for its capability in motion sensing, has been widely applied in contactless respiration monitoring [7-10], offering the advantages of bringing no burden to subjects. The chest movements captured by radar exhibits strong correlation with the respiratory effort from PSG [11], making it possible for radar to detect SAE. Although radar is unable to capture EEG signals, different sleep stages are associated with distinct patterns in vital signs such as breathing rate, heart rate, and body movement [12]. For example, individuals typically show greater variability in respiratory rate

Corresponding author: Gang Li (gangli@tsinghua.edu.cn)
This work was supported by National Natural Science Foundation of China under Grants 61925106.
Wei Wang, Xiang Zhao and Gang Li are with the Department of Electronic Engineering, Tsinghua University, Beijing 100084, China.
Zhaoxi Chen, Wenyu Zhang and Zetao Wang are with the Beijing Qinglei Technology Co. Ltd., Beijing 100089, China.
Chenyang Li, Jian Guan and Shankai Yin are with the Department of Otolaryngology-Head and Neck Surgery & Center of Sleep Medicine, Shanghai JiaoTong University school of medicine Affiliated Sixth People's Hospital, Shanghai Key Laboratory of Sleep Disordered Breathing, Shanghai 200233, China.

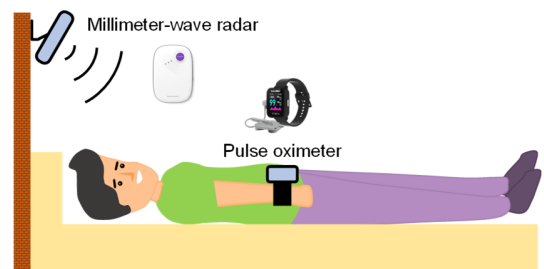

Fig. 1. Monitoring scene of ROSA

and higher body movement frequency during wakefulness. During deep sleep, respiratory rate becomes more stable, and body movements are significantly reduced. Radar can measure all these vital signs, allowing it to classify sleep stages.

In recent years, many studies have employed radar technology for sleep monitoring, including SAE detection [11, 13-19] and sleep stage classification [20-26]. Although the classification performance of radar declines as the granularity of sleep stages increases [22-26], it performs well in distinguishing between wakefulness and sleep [20, 21]. This is sufficient to meet the requirements of our study to accurately estimate TST and calculate AHI. Researches on SAE detection can be categorized into handcrafted feature-based methods [11, 14, 19, 27] and deep learning methods [15-17]. Handcrafted feature-based methods are derived from human experience, whose performance is influenced by individual variability. Deep learning methods can learn hierarchical representations from raw data, which demonstrate good performance on large-scale datasets. Existing deep learning methods for SAE detection typically divide the preprocessed data (temporal signal, spectrogram, etc.) into segments with fixed length, and then classify divided segments as either normal or abnormal [15-17]. These methods typically necessitate a complex post-processing workflow to translate segment-level detection into event-level detection, as the estimation of AHI involves calculating the number of SAE. Additionally, segment division strategies require careful consideration [28], because segments that are either too long or too short can affect the detection of abnormal events. The approach of outputting bounding boxes in object detection offers potential solutions to these challenges. Related models have been modified and applied to SAE detection using airflow [29], abdominal breathing [30], and similar inputs [31], but their application to radar-based SAE detection remains unexplored. Faster R-CNN is a widely utilized model in object detection, which directly outputs bounding boxes along with their associated categories [32]. A variant of this architecture has also been introduced to facilitate the temporal localization of specific actions within video sequence [33] . Both action localization and SAE detection can be conceptualized as one-dimensional (1D) object detection tasks. Consequently, the methods outlined in [33] provide significant insights for radar-based SAE detection.

While radar-based SAE detection performs well in controlled lab settings, it is still prone to errors when affected by body movements or environmental interference. $SpO_2$ can be obtained by a low-burden pulse oximeter. Recent studies have shown that $SpO_2$ provides useful information for SAE detection [34-37], but using it solely for OSAHS diagnosis also has certain limitations [38]. For example, changes in hemoglobin levels can artificially raise or lower $SpO_2$, which is unrelated to SAE. Moreover, some sleep-related breathing disorders which cause no oxygen desaturation (OD) may be missed when using $SpO_2$ alone.

In this study, we find that fusing radar and oximeter data can achieve better performance in SAE detection compared to using either data source alone. Therefore, we introduce a method using millimeter-wave radar and pulse oximeter for the diagnosis of OSAHS, called ROSA. It includes a deep learning network designed for radar-based SAE detection utilizing an R-CNN architecture (RASA R-CNN), a network designed for radar-based sleep stage classification (RassNet), and a decision-level fusion strategy to integrate $SpO_2$ information from oximeter data. Expanding on our previous work [39], this study further presents a soft-fusion strategy and integrates the sleep staging network in ROSA. This enhancement allows ROSA to automatically estimate AHI and diagnose OSAHS using only radar and pulse oximeter. We also compare the performance of our method with previous studies. Experimental results demonstrate that ROSA outperforms previous methods in AHI estimation and OASHS diagnosis. While achieving high agreement with PSG, ROSA significantly reduces the burden on subjects, making it more conducive to widespread adoption. Main contributions of this study are follows:

- We develop a radar-based detection network using R-CNN structure, called RASA R-CNN, which takes the pre-processed spectrograms of radar signals as input and outputs the temporal localization of SAE directly.
- We design a deep learning network for radar-based sleep stage classification, which facilitates the calculation of TST in AHI estimation.
- We present a soft-fusion strategy to integrate $SpO_2$ information, which can eliminate the false alarm of SAE detected by RASA R-CNN and improve the performance of SAE detection.
- We collect real-world data in a hospital setting to validate the agreement between ROSA and PSG in AHI estimation and OSAHS diagnosis, and demonstrate the advantages of ROSA compared to previous studies.

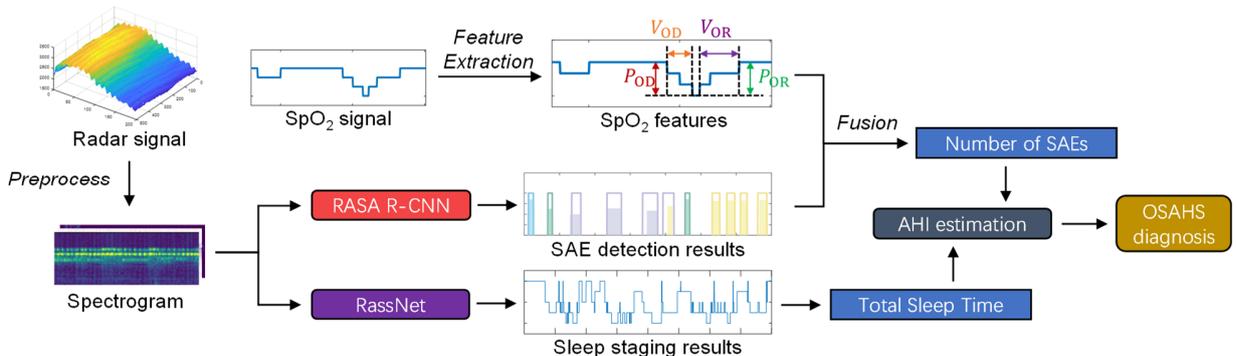

Fig. 2. Overall process of ROSA

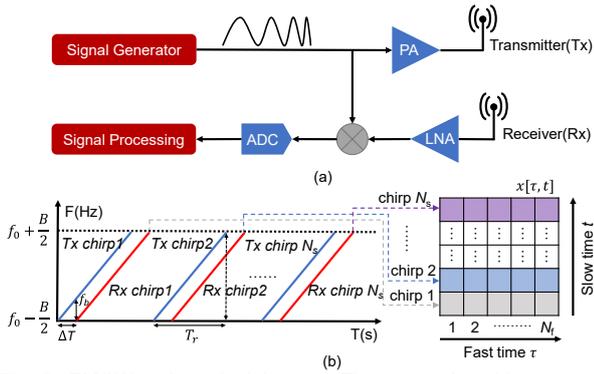

Fig. 3. FMCW radar principles. (a) The general architecture of FMCW radar; (b) The time-frequency response of Tx/Rx signal and a visual representation of $x[\tau,t]$.

## II. METHODOLOGY

The radar used in this study is placed above the head of the bed (Fig. 1), whose LOS faces the chest of the subject. Subjects are additionally instructed to wear a fingertip pulse oximeter to simultaneously record $SpO_2$ signals. The overall process of ROSA is shown in Fig. 2.

### A. Radar Signal Pre-processing

During pre-processing, the received signals from frequency modulated continuous wave (FMCW) radar are converted into three distinct spectrograms, each capturing a specific physical characteristic. These spectrograms show the spatial and temporal variations of the corresponding physical parameters.

The general architecture of FMCW radar is shown in Fig. 3(a). It periodically transmits a frequency modulated sinusoid signal whose frequency is swept linearly from $f_0-B/2$ to $f_0+B/2$ with a slope of $K$ and a duration of $T_r$, as shown in Fig. 3(b). The received signals are mixed with the transmitted signals to obtain the beat signals $x[\tau,t]$ [10]. Assuming that the distance from radar to the subject is $R_0$ and $d[t]$ represents the micro-movements of the body like breathing, the beat signal $x[\tau,t]$ can be written as:

$$x[\tau,t] \approx A_b \exp\left(2\pi j\left(2K\frac{R_0}{c}\tau + 2\frac{R_0+d[t]}{\lambda_0}\right)\right), \quad (1)$$

where $A_b$ is the amplitude of the beat signal, $\lambda_0 = c/f_0$ is the wavelength corresponding to the carrier frequency, $\tau$ represents the fast time (timestamps within a chirp) and $t$ represents the slow time (timestamps across chirps). The windowed fast Fourier transform is applied along the fast time dimension of $x[\tau,t]$ to obtain the range-slow-time matrix $R[r,t]$, where $r$ denotes the range.

Then we conduct a series of filtering operations along the slow-time dimension of $R[r,t]$. The high-pass filtering technique, employing a cut-off frequency of $f_c$=5Hz, is utilized to extract signals that capture human body movements. To isolate respiration-related signals, the band-pass filtering technique with a passband ranging from 0.1Hz to 5Hz is utilized. The first spectrogram $x_M[r,t]$, representing the power associated with the body movement, is derived from the high-frequency signals. The second spectrogram $x_B[r,t]$, representing the power of breathing, is derived from the respiration-related signals. The chest movements associated with respiration can introduce doppler frequency shifts in the received signals. We conduct doppler analysis on the respiration-related signals to derive the two-dimensional distribution of the principal component of doppler frequency over range and time, denoted as the third spectrogram $x_D[r,t]$. Fig. 4 illustrates an example of three pre-processed spectrograms aligned with airflow, respiratory effort and $SpO_2$ signals from PSG. When OA occurs, the PSG signals indicate a cessation of airflow and a reduction in respiratory effort. Correspondingly, the reduction of breathing power and doppler in the pre-processed spectrograms can be observed at the same temporal location. The three pre-processed spectrograms are concatenated along the channel dimension to construct a new spectrogram with three channels, which serves as the input for the proposed deep learning model.

### B. Deep Learning Model Architecture

#### 1) Sleep Staging Network

The sleep staging network, RassNet, takes the concatenated spectrogram as input and outputs the logits of sleep stage, i.e. Wake (W), REM (R), N1, N2 and N3, as shown in Fig. 5(a). First, ResNet18 is used as the backbone in RassNet to extract features (Fig. 5(b)). Second, considering the periodicity of sleep structure, we utilize a long short-term memory (LSTM) layer to learn the temporal dependencies between different sleep stages. Instead of employing direct summation or concatenation, we also adopt the attentional feature fusion (AFF) [40] module for skip connections to facilitate the more effective fusion of features in different levels. The LSTM layer outputs the logits of different sleep stages. We denote the logits at the $n$-th time step as $\mathbf{y}_n = [y_n^W, y_n^R, y_n^{N1}, y_n^{N2}, y_n^{N3}]$.

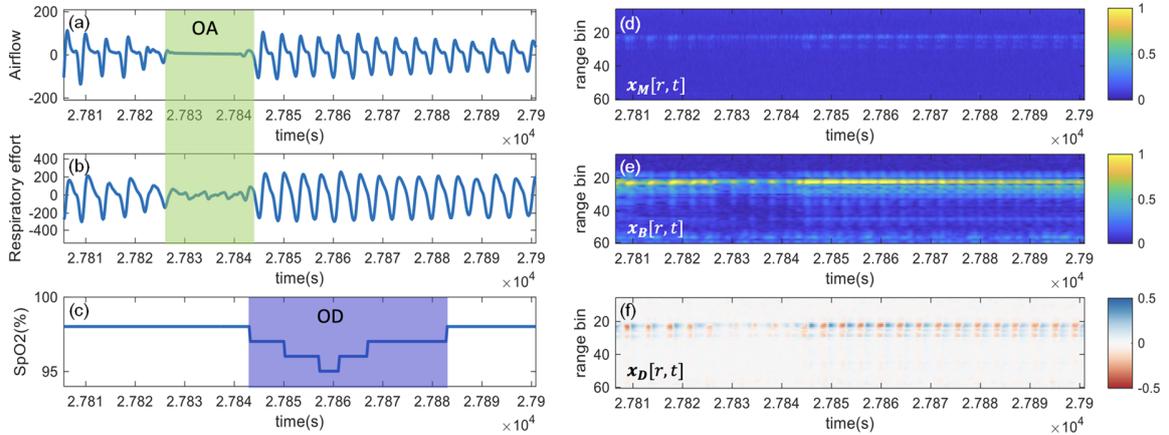

Fig. 4. Pre-processed spectrograms aligned with physiological signals. (a)-(c) Airflow, respiratory effort and $SpO_2$ signal from PSG; (d)-(f) The spectrograms of body movement power, breathing power and breathing doppler obtained by radar signal pre-processing.

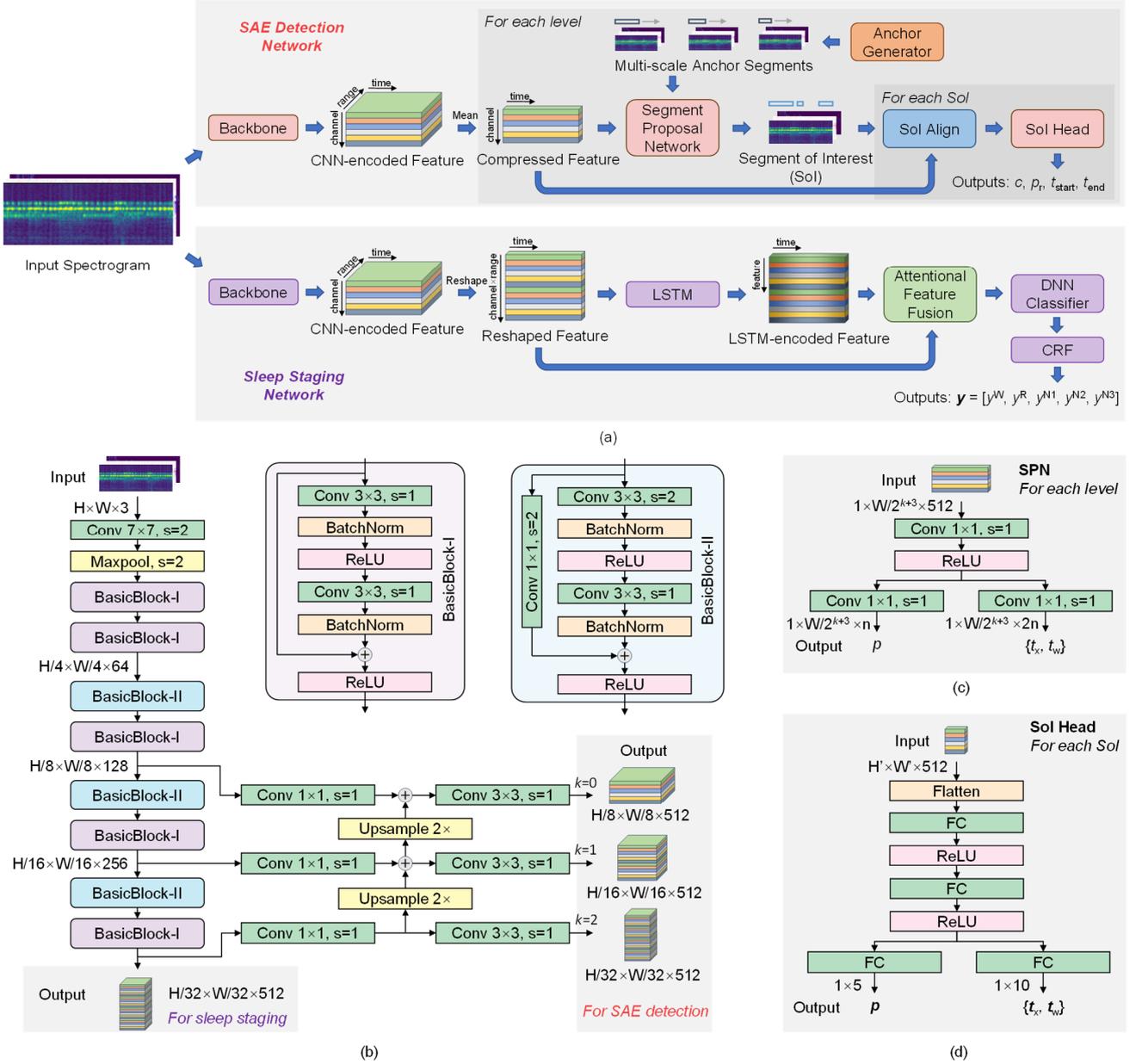

Fig. 5. Network architecture. (a) Overall architecture in ROSA; (b) Backbone architecture; (c) SPN architecture; (d) SoI Head architecture.

According to medical research, the transition of sleep stage always follows a certain rule [41]. We also find that the results predicted by the aforementioned features exhibit certain inconsistencies, characterized by irregular transitions that deviate from the rule governing sleep stage transitions. To mitigate this issue, we add a conditional random field (CRF) layer [42] after LSTM, which can learn the transition score between different sleep stages. At the inference stage, the CRF layer will perform Viterbi decoding of the logits from the LSTM layer to obtain the final results $\mathbf{y}'_n$.

*2) SAE Detection Network*

Numerous deep learning models have been proposed for the task of SAE detection based on respiratory signals [17, 28, 43]. These works first divide respiratory signals or spectrograms into segments using a sliding window with fixed temporal length. Each segment is labelled through a certain mechanism, e.g., majority voting. With the utilization of segment-based methods, a long-duration event may be divided into multiple segments. Conversely, several short-duration events may be divided into the same segment, presenting difficulties in assigning appropriate labels. We also find it challenging to accurately quantify the number of events under the aforementioned conditions, even with complex post-processing procedures. Therefore, these methods have limitations for the calculation of AHI, which is the key factor in OSAHS diagnosis. To mitigate this issue, we propose an event-level detection network architecture, RASA R-CNN, inspired by the classical architectures in the field of object detection.

The SAE detection network, RASA R-CNN, follows the detection paradigm of Faster R-CNN [32], as shown in Fig. 5(a). Object detection is committed to detect two-dimensional (2D) spatial regions containing specific objects. In this study, the goal of SAE detection is to detect 1D temporal segments containing specific events, which can be regarded as 1D object detection. Therefore, we find that the detection paradigm of Faster R-CNN can be effectively applied to the task of SAE detection with certain modifications.

RASA R-CNN takes the concatenated spectrogram as input and outputs the parameters of all detected segments. Each detected segment is characterized by four parameters: $c$, $p_r$, $t_{start}$ and $t_{end}$, denoting the category, score, start time and end time, respectively. It contains two stages similar to Faster R-CNN. First, we utilize a ResNet18 with feature pyramid structure as the backbone, and outputs the 2D feature maps of different levels ($k$=0,1,2) for SAE detection. High-level features are effective for detecting long-duration events, while low-level features are more sensitive to short-duration events. For each level, the 2D feature map is compressed along the range dimension and subsequently inputted into a segment proposal network (SPN), which is a 1D variant of the region proposal network (RPN) [32]. Anchor segments of various scales are generated as candidate segments at each time step, which are then used by SPN to propose segments of interests (SoIs) for the next stage and regresses their boundaries (Fig. 5(a)). Each SoI is characterized by its score $p$ and regression coefficients $\{t_x, t_w\}$. The temporal localization of SoI can be calculated as:

$$x = x_a + t_x w_a, w = w_a \exp t_w, \quad (2)$$

where $x$ and $w$ denote the center and width of SoI, $x_a$ and $w_a$ denote the center and width of the anchor. Second, SoIs are aligned using a 1D variant of the RoIAlign layer in Mask R-CNN [44]. The SoI head is used to predict the category of SoIs and further regress their boundaries. As shown in Fig. 5(d), the SoI head outputs the scores of various types of SAEs $\boldsymbol{p}$=[$p_0$, $p_1$, $p_2$, $p_3$, $p_4$] and corresponding regression coefficients $\{\boldsymbol{t_x}$=[$t_{x,0}$, $t_{x,1}$, $t_{x,2}$, $t_{x,3}$, $t_{x,4}$], $\boldsymbol{t_w}$=[$t_{w,0}$, $t_{w,1}$, $t_{w,2}$, $t_{w,3}$, $t_{w,4}$]$\}$, where 0~4 denote normal, CA, OA, MA and HP. Finally, the SAE detection module outputs all detected segments represented by parameters introduced above, which can be calculated as:

$$c = \arg\max_i p_i, i = 0,1,2,3,4 \quad (3)$$

$$p_r = p_c, \quad (4)$$

$$t_{start} = x + t_{x,c} w - \frac{w \exp t_{w,c}}{2}, \quad (5)$$

$$t_{end} = x + t_{x,c} w + \frac{w \exp t_{w,c}}{2}. \quad (6)$$

*3) Loss Function*

The loss function of RASA R-CNN is a 1D variant of Faster R-CNN [32], which is a combination of classification loss and regression loss.

The loss function of RassNet, denoted as $\mathcal{L}_{stage}$, combines of classification loss, change loss, duration loss and transition loss. The classification loss is utilized by focal loss [45] to recognize hard samples better. It can be written as

$$\mathcal{L}_{Focal} = -\frac{1}{N} \sum_{n=0}^{N-1} \omega_{s_n^*} \left(1 - p_n^{s_n^*}\right)^2 \log p_n^{s_n^*}, \quad (7)$$

$$p_n^{s_n^*} = \frac{\exp\left(y_n^{s_n^*}\right)}{\sum_{k=W,R,N1,N2,N3} \exp(y_n^k)}, \quad (8)$$

where $N$ is the number of time steps, $s_n^* \in [W, R, N1, N2, N3]$ is the true sleep stage at the $n$-th time step and $\omega_{s_n^*}$ is the weighting factor.

To punish short-duration sleep stages predicted by radar, we propose a change loss and adopt the duration loss in [41].

$$\mathcal{L}_{change} = \frac{1}{N-1} \sum_{n=1}^{N-1} \|y_n - y_{n-1}\|^2, \quad (9)$$

$$\mathcal{L}_{duration} = \sum_{n=1}^{N-1} \sum_i \text{ReLU}(T^i - C_{n-1}^i)(1 - p_n^i), \quad (10)$$

$$C_n^i = \begin{cases} 0, & n = 0 \\ y_{n-1}^i(C_{n-1}^i + d) + (1 - C_{n-1}^i)d, & n \neq 0 \end{cases}, \quad (11)$$

where $T^i$ is the minimum duration of stage-$i$, $d$ is the unit duration and $C_n^i$ is the expected duration of stage-$i$ till the $n$-th time step.

The transition loss is implemented by the CRF loss, which considers the logits outputted from LSTM and the transition scores between different sleep stages. For a sleep stage sequence $\boldsymbol{s} = [s_0, s_1, \dots, s_{N-1}]$, its score is defined as:

$$g(\boldsymbol{s}) = \sum_{n=0}^{N-2} A_{s_n, s_{n+1}} + \sum_{n=0}^{N} y_n^{s_n}, \quad (11)$$

where $A_{i,j}$ denotes the transition score from stage-$i$ to stage-$j$. Then we can calculate the probability for the sequence $\boldsymbol{s}$ by a SoftMax function:

$$p(\boldsymbol{s}) = \frac{\exp g(\boldsymbol{s})}{\sum_{\tilde{\boldsymbol{s}}} \exp g(\tilde{\boldsymbol{s}})}, \quad (12)$$

where $\tilde{\boldsymbol{s}}$ denotes all possible sequences. In the training stage, we want to maximize the probability of the true sleep stage sequence $\boldsymbol{s}^*$. Therefore, the CRF loss can be written as:

$$\mathcal{L}_{CRF} = -\log p(\boldsymbol{s}) = \log \sum_{\tilde{\boldsymbol{s}}} \exp g(\tilde{\boldsymbol{s}}) - g(\boldsymbol{s}). \quad (13)$$

In summary, the loss function of RassNet can be written as:
$$\mathcal{L}_{stage} = \alpha \mathcal{L}_{Focal} + \beta \mathcal{L}_{change} + \gamma \mathcal{L}_{duration} + \eta \mathcal{L}_{CRF}, \quad (14)$$
where $\alpha, \beta, \gamma, \eta$ are weighting coefficients.

*C. Fusion of Radar and Oxygen Saturation Data*

SAE typically causes recurrent oxygen desaturation (OD) [46]. To fuse information obtained by radar and pulse oximeter, we first perform feature extraction on the SpO$_2$ signals within 1 minute following each segment detected by RASA R-CNN. Specifically, we extract the decrease percentage of the first OD exceeding 3%, denoted as $P_{OD}$, along with the subsequent oxygen saturation rise (OR) percentage $P_{OR}$ (shown in Fig. 6). If there is no OD greater than 3%, we extract the percentage of the maximum OD. We enhance the robustness of feature extraction by disregarding short-duration fluctuations (<10s) in blood oxygen levels. We further evaluate the reliability of radar-detected segments based on the SpO$_2$ features, achieving the fusion of radar and oxygen saturation. In our real-world dataset, occasional artifacts are observed in the SpO$_2$ signals, characterized by SpO$_2$ values of 0 or 255. Data points corresponding to these instances of artifacts are excluded from the fusion process.

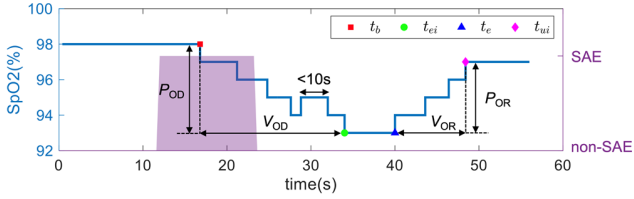

Fig. 6. An example of SpO$_2$ features. The features used in Algorithm 1 are annotated in the figure. A fluctuation in blood oxygen levels characterized by duration less than 10s and amplitude equal to 1% is disregarded, demonstrating the robustness of our algorithm.

The soft-fusion strategy aims to assign a SpO$_2$-based score to the radar-detected segments through a light-weight neural network, denoted as $p_s \in [0,1]$. The scores provided by the soft fusion exhibit heightened sensitivity to variability in SpO$_2$ features. Therefore, it can potentially improve the accuracy of fusion results and make effective use of SpO$_2$ information.

To train the neural network for soft fusion, we need to construct a sub-dataset $\mathcal{D}$ consisting of vectors that represent SpO$_2$ features, each being annotated with the corresponding type of SAE or non-SAE. Specifically, we first extract the above SpO$_2$ features for all true SAE. These feature vectors are annotated with the corresponding type of SAE. Then, we randomly select a certain number of SpO$_2$ segments without SAE and their feature vectors are annotated as non-SAE.

In the training stage, we construct the aforementioned sub-dataset on the training set of the whole dataset. The cross-entropy loss function is employed to train our light-weight network, consisting of three fully connected layers. In the inference stage, we extract SpO$_2$ features of for each radar-detected segment. The feature vector is then fed into the trained light-weight network to obtain the SpO$_2$-based score $p_s$. We then summarize the scores from RASA R-CNN and light-weight network with the weighting coefficients. The detailed process of the soft-fusion strategy is presented in Algorithm 1.

## III. EXPERIMENTAL RESULTS

### A. Data Collection

We used a FMCW radar system based on Infineon BGT60TR13C and a ChoiceMMed MD300W628 pulse oximeter to simultaneously monitor subjects who undergo PSG during sleep (Fig. 1). Infineon-BGT60TR13C is a 60GHz Antenna-In-Package (AiP) radar equipped with one transmitter and three receivers. We used a single receiver in our study. The radar parameters are shown in Table I. PSG provides the ground truth of SAE and sleep stages. All PSG data were annotated by sleep technologists, according to the AASM manual for the scoring of sleep and associated events [47].

The dataset containing overnight recordings of 100 subjects

---

**Algorithm 1** Soft Fusion based on Light-weight Network

**Input:** Scores of detected segments $\boldsymbol{p}_r = \{p_r^i\}_{0 \leq i \leq N-1}$; SpO$_2$ features $[\boldsymbol{P}_{OD}, \boldsymbol{P}_{OR}, \boldsymbol{V}_{OD}, \boldsymbol{V}_{OR}] = \{P_{OD}^i, P_{OR}^i, V_{OD}^i, V_{OR}^i\}_{0 \leq i \leq N-1}$, where $N$ is the number of radar-detected segments; dataset $\mathcal{D}$ containing SpO$_2$ features corresponding to different types of SAE and non-SAE.

**Output:** Scores after soft fusion $\boldsymbol{p}_f = \{p_f^i\}_{0 \leq i \leq N-1}$.

**1 Training:**
Train the light-weight network $\mathcal{M}$ on dataset $\mathcal{D}$.
**2 Inference:**
Use the trained network $\mathcal{M}$ to assign a SpO$_2$-based score $p_s^i$ to each detected segment.
**for** $i = 0 : N-1$
  $p_s^i = \mathcal{M}(P_{OD}^i, P_{OR}^i, V_{OD}^i, V_{OR}^i)$
**end**
**3 Fusion:**
weighting coefficient $\omega$;
**for** $i = 0 : N-1$
  $p_f^i = \omega p_r^i + (1-\omega) p_s^i$
**end**
**return** $\boldsymbol{p}_f$

---

(>800 hours) for this study were collected from the Shanghai JiaoTong University school of medicine Affiliated Sixth People's Hospital. This study was conducted in accordance with the Declaration of Helsinki and the study protocol was approved by the Ethics Committee of Shanghai JiaoTong University Affiliated Sixth People's Hospital (2023-030-[1]). The study was registered at the United States Clinical Trial Registry (No. NCT06038006). All subjects provided informed consent. The basic information of the subjects is shown in Table II. The subjects were divided into four groups based on the value of PSG-derived AHI: healthy group (AHI<5, $n$=27), mild OSA group (5≤AHI<15, $n$=32), moderate OSA group (15≤AHI<30, $n$=16) and severe OSA group (AHI≥30, $n$=34).

### B. Implementation Details

The proposed RASA R-CNN and RassNet is implemented on Pytorch. We train the RASA R-CNN by Adam optimizer for 1500 epochs. We train the RassNet by Adam optimizer for 1000 epochs. The learning rate is set with reference to the cosine annealing learning rate. The weighted cross entropy loss is used to deal with data imbalance. In our experiment, we applied 4-

Table I
CONFIGURATIONS OF FMCW RADAR

| Parameter | Configuration |
|---|---|
| Carrier frequency $f_0$ | 60 GHz |
| Sweep bandwidth $B$ | 3 GHz |
| Frame rate $F$ | 250 Hz |
| Samples per chirp $n$ | 256 |

Table II
BASIC INFORMATION OF SUBJECTS

| Group | Num | TST (h) | AHI (events/h) |
|---|---|---|---|
| Healthy | 27 | 7.05±1.70 | 2.34±1.04 |
| Mild | 32 | 6.73±1.56 | 8.09±2.19 |
| Moderate | 16 | 7.38±1.22 | 21.90±3.94 |
| Severe | 25 | 7.31±1.01 | 57.22±18.10 |

## Table III
### SLEEP STAGING RESULTS UNDER DIFFERENT TASKS AND LOSS FUNCTIONS

| Training strategy | Loss function | | | | WS | | WRLD | | WRNN | |
|---|---|---|---|---|---|---|---|---|---|---|
| | $\mathcal{L}_{\text{Focal}}$ | $\mathcal{L}_{\text{duration}}$ | $\mathcal{L}_{\text{change}}$ | $\mathcal{L}_{\text{CRF}}$ | Acc (%) | $\kappa$ | Acc (%) | $\kappa$ | Acc (%) | $\kappa$ |
| One-stage | ✓ | ✗ | ✗ | ✗ | 93.69 | 0.8393 | 78.89 | 0.7005 | 71.03 | 0.6318 |
| | ✓ | ✓ | ✗ | ✗ | 94.02 | 0.8471 | 79.93 | 0.7107 | 72.78 | 0.6510 |
| | ✓ | ✓ | ✓ | ✗ | **94.43** | **0.8594** | 79.98 | 0.7165 | 72.55 | 0.6489 |
| | ✓ | ✓ | ✓ | ✓ | 93.91 | 0.8467 | 79.82 | 0.7105 | 73.10 | 0.6509 |
| Two-stage | ✓ | ✓ | ✓ | ✓ | 94.16 | 0.8546 | **80.73** | **0.7233** | **74.72** | **0.6688** |

fold cross-validation and reported the performance on the test set. When using two-stage training strategy, we set $\eta=0$ in the first 500 epochs, and set $\eta=1$ in the last 500 epochs.

### C. Evaluating Metrics
#### 1) Sleep Staging Evaluation
The sleep staging results obtained by RassNet are compared with those from PSG. We adopted accuracy (Acc) and Cohen's kappa ($\kappa$) to evaluate the overall sleep staging performance. Intraclass correlation coefficient (ICC) [48] is used as the metric to evaluate the agreement between the estimated and PSG-derived TST.

#### 2) SAE Detection Evaluation
The SAE detected by ROSA are compared with those annotated by sleep technologists to evaluate the performance of our method. The average precision (AP), a widely used metric in object detection, is employed as the primary metric to evaluate the SAE detection performance. In particular, we used $AP_{0.5}$ in this study, which refers to the AP calculated at an intersection over union (IoU) threshold of 0.5.

AHI serves as a crucial metric for diagnosing and quantifying the severity of OSAHS. It delineates the average number of SAE per hour of sleep. The definition of AHI is

$$\text{AHI} = \frac{N_{\text{apnea}} + N_{\text{hypopnea}}}{\text{TST}}, \quad (15)$$

where $N_{\text{apnea}}$ is the number of apnea events, $N_{\text{hypopnea}}$ is the number of hypopnea events, and TST, measured in hours, is the total sleep time calculated according to the sleep staging results. ICC is also used as the metric to compare the agreement between the AHI estimated from ROSA and that provided by PSG. To evaluate the diagnostic performance at different AHI thresholds, we adopted sensitivity (Se), specificity (Sp), accuracy and Cohen's kappa in this study.

### D. Performance of Sleep Staging
#### 1) Results of Different Sleep Staging Tasks
Table III presents the detailed results of sleep staging by the proposed method. The performance of the RassNet is evaluated under different tasks, including Wake-Sleep (WS) classification, Wake-REM-Light-Deep (WRLD) classification and Wake-REM-N1-N2-N3 (WRNN) classification. The results show that RassNet has good performance in all sleep staging tasks. Particularly, the accuracy and Kappa coefficient in WS classification can achieve 94.43% and 0.8594, demonstrating sufficient reliability for TST estimation. We calculate the duration of the Sleep stage as the estimated value of TST. The comparison of the estimated and PSG-derived TST is shown in Fig. 7.

Ablation experiments with various loss functions are also performed to evaluate the impact of each on the training process, as shown in Table III. For the CRF loss, we specifically design a two-stage training strategy to maximize its effectiveness. The two-stage training strategy first uses the classification loss, change loss and duration loss to train the RassNet except the CRF layer. Based on the pretrained parameters, the strategy then uses the transition loss to train the CRF layer and fine-tune other layers. Results obtained by ablation experiments show that the proposed change loss can significantly improve the performance of WS classification. Combined with the two-stage training strategy, the transition loss is highly beneficial for the detailed sleep stage classification, showing the best accuracies of 80.73% and 74.72% in WRLD and WRNN classification. However, the transition loss provides almost no enhancement in WS classification. This may be due to its primary objective of guiding the network to learn transition patterns between different sleep stages, while such patterns are not clear between wakefulness and sleep.

#### 2) Comparison with Radar-based Sleep Staging of Previous Studies
Table IV presents the performance of recent studies on radar-based sleep staging on their respective datasets. Our study

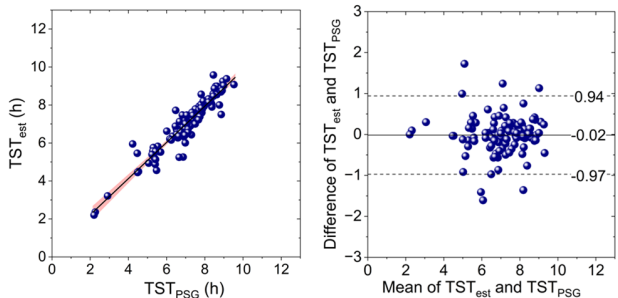

Fig. 7. The scatter plot and Bland-Altman analysis of the estimated and PSG-derived TST.

### Table IV
#### COMPARISON OF SLEEP STAGING PERFORMANCE WITH PREVIOUS STUDIES

| Task | Study | RF Sensors | N | Performance | |
|---|---|---|---|---|---|
| | | | | Acc (%) | $\kappa$ |
| WS | Pallesen et al. [20] | IR-UWB | 14 | 93.1 | 0.670 |
| | Heglum et al. [21] | IR-UWB | 62 | 94.8 | 0.869 |
| | | | 28 | 80.9 | 0.448 |
| | Our study | FMCW | 100 | 94.4 | 0.859 |
| WR-LD | Tataraidze et al. [22] | FMCW | 32 | 63.5 | 0.490 |
| | Toften et al. [23] | IR-UWB | 71 | 76.0 | 0.630 |
| | Zhai et al. [24] | CW | 33 | 79.2 | 0.679 |
| | Zhao et al. [25] | RF | 25 | 79.8 | 0.700 |
| | Kwon et al. [26] | IR-UWB | 51 | 82.6 | 0.730 |
| | Our study | FMCW | 100 | 80.7 | 0.723 |

WS, Wake-Sleep; WRLD, Wake-REM-Light-Deep; RF, Radio frequency; FMCW, Frequency-modulated continuous-wave; IR-UWB, Impulse-radio ultra-wideband; N, Number of subjects.

## Table V
### COMPARISON OF SAE DETECTION PERFORMANCE ACROSS DIFFERENT SENSORS AND METHODS

| Sensor | Method | $AP_{0.5}$ (%) | ICC | D.T. (#/h) | Se (%) | Sp (%) | Acc (%) | $\kappa$ |
|---|---|---|---|---|---|---|---|---|
| Oxi-meter | OxiNet [34] | / | 0.7512 | 5 | 61.64 | 100.00 | 72.00 | 0.4646 |
| | | | | 15 | 78.05 | 98.31 | 90.00 | 0.7870 |
| | | | | 30 | 72.00 | 97.33 | 91.00 | 0.7429 |
| | $ODI_3$ | / | 0.9069 | 5 | 72.60 | 100.00 | 80.00 | 0.5886 |
| | | | | 15 | 85.37 | 98.31 | 93.00 | 0.8526 |
| | | | | 30 | 92.00 | 98.67 | 97.00 | 0.9189 |
| Radar | RASA R-CNN | 68.05 | 0.9740 | 5 | 87.67 | 85.19 | 87.00 | 0.6884 |
| | | | | 15 | 92.68 | 96.61 | 95.00 | 0.8963 |
| | | | | 30 | 96.00 | 93.33 | 94.00 | 0.8481 |
| Fusion | ROSA | 74.23 | 0.9870 | 5 | 90.41 | 88.89 | 90.00 | 0.7576 |
| | | | | 15 | 90.24 | 96.61 | 96.00 | 0.9161 |
| | | | | 30 | 96.00 | 98.67 | 98.00 | 0.9467 |

ICC: Intraclass correlation coefficient; $ODI_3$: 3% oxygen desaturation index; D.T.: diagnostics thresholds of AHI.

shows better performance than most related studies, expect [21] and [26], which were conducted only on healthy subjects. It is worth noting that healthy subjects have more regular sleep structure than OSA patients, making sleep staging on them less challenging. Additionally, the larger sample size in our study increased the confidence of our results.

### E. Performance of SAE Detection
#### 1) $SpO_2$-based SAE Detection

We first test the performance of $SpO_2$-based SAE detection. The OxiNet proposed in [34] is used to estimate the AHI. The pretrained model which training on sufficient public dataset is directly used in our study. For each subject, the pretrained OxiNet takes the $SpO_2$ signal throughout the night as input, and outputs the estimated AHI value. We also use the 3% oxygen desaturation index ($ODI_3$) as an estimation for AHI [46]. The detailed results of $SpO_2$-based SAE detection are presented in Table V. The experiment results demonstrate the efficacy of using $ODI_3$ to estimate the AHI (ICC=0.9069). The performance of OxiNet on our data is noticeably worse than what was reported in [34]. This difference may be attributed to variations in the oximeters used across different studies. Oxygen desaturation typically follows a sleep apnea event, and the delay time is often irregular. While the $SpO_2$ signal provides valuable information for SAE detection, it is challenging to precisely localize these events based solely on this signal. As a result, it is not possible to calculate the $AP_{0.5}$ for methods that using oximeter data only.

#### 2) Radar-based SAE Detection

We then test the performance of radar-based SAE detection, Radar-based SAE detection is implemented by RASA R-CNN without the fusion of radar and $SpO_2$ signals. It detects SAE primarily by analyzing changes in breathing patterns reflected in the preprocessed spectrograms. The results in Table V show the good performance of RASA R-CNN in SAE detection ($AP_{0.5}$=68.05%). The detection results in Fig. 8 demonstrate that RASA R-CNN can not only detect the occurrence of SAE but also accurately identify their temporal locations.

#### 3) Fusion-based SAE Detection

The fusion of radar and $SpO_2$ signal serves to mitigate false alarms and enhance the confidence of accurate event detection. Table V presents the detailed results of SAE detection using the presented fusion strategy. Fig. 8 presents an example of SAE detection during a subject's sleep. The experimental results demonstrate that the overall detection performance exhibits superiority with the fusion of radar and $SpO_2$ signal using ROSA, resulting in an improvement of 6.18% in $AP_{0.5}$. By leveraging the learned scores, the soft-fusion strategy effectively integrates the information from radar and $SpO_2$ signals, thereby improving the robustness and accuracy in SAE detection. The significant SAE detection result provides support for subsequent estimation of the AHI and the diagnosis of OSAHS.

#### 4) Classification of Various Types of SAE

As introduced in Section I, SAE can be divided into CA, OA, MA and HP. HP refers to a reduction in airflow during sleep without a complete cessation of breathing, accompanied by decreased but still present respiratory effort. CA occurs due to a disruption in the central control of breathing, leading to a temporary cessation of respiratory effort and a complete absence of airflow. In contrast, OA is caused by airway obstruction, resulting in a significant reduction in airflow while respiratory effort continues. Accurate classification of SAE is essential for clinicians to better assess the patient's condition and develop effective treatment plans.

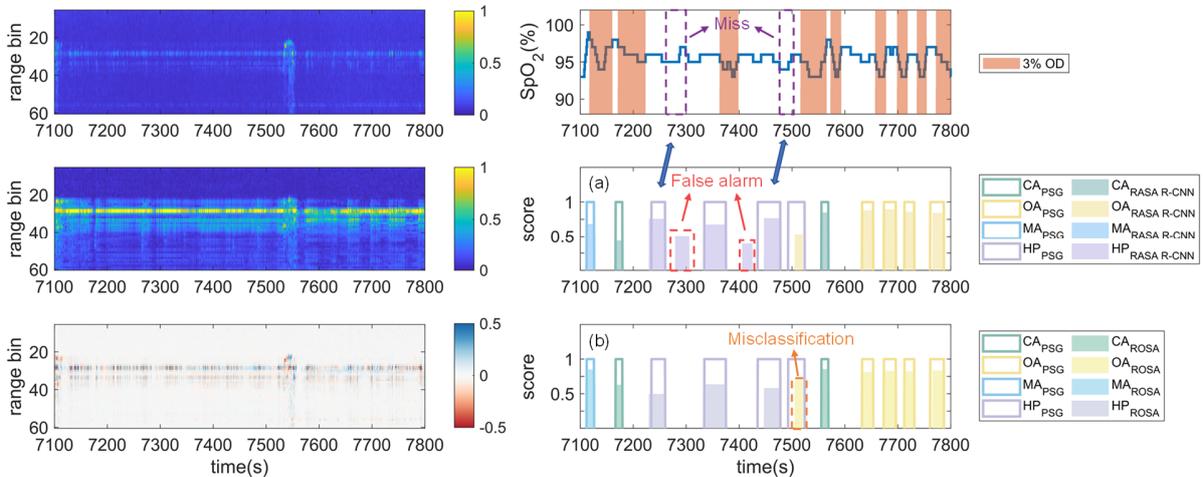

Fig. 8. Radar spectrogram, $SpO_2$ signal and corresponding SAE detection results. (a) RASA R-CNN; (b) ROSA.

Table VI
DETECTION PERFORMANCE OF EACH TYPE OF SAE

| Method | AP$_{0.5}$ (%) | | | |
|---|---|---|---|---|
| | CA | OA | MA | HP |
| RASA R-CNN | **33.81** | 55.84 | **23.71** | 19.05 |
| ROSA | 27.25 | **57.06** | 20.32 | **20.19** |

CA: central apnea; OA: obstructive apnea; MA: mixed apnea; HP: hypopnea;

Table VI presents the detection performance of each type of SAE obtained from both RASA R-CNN and ROSA. As Table VI shows, all methods can distinguish between different types of SAE. The detection performance of OA is superior to that of all other types of SAE, achieving an AP$_{0.5}$ of 57.06%. OA is the most common type of SAE. It has the largest sample size and distinct characteristics, which makes it easier to recognize. Even though HP frequently occurs, its subtler reduction in airflow makes it less distinguishable compared to other apnea events. Therefore, the detection performance of HP is the worst among all types of SAE, with an AP$_{0.5}$ of only 20.19%.

Table V and Table VI also indicate that while the fusion of radar and SpO$_2$ signals can enhance overall detection performance, it does not improve the detection accuracy of all types of SAEs. For instance, the fusion method leads to decreased AP values for CA and MA. CA is a type of SAE which may not result in OD. Therefore, SpO$_2$ signal can only provide limited useful information for CA detection, and may even mislead the detection of this type. MA, defined as combination of CA and OA, is similarly affected by this limitation. It shows that while ROSA offers benefits for SAE detection, the impact may vary depending on the characteristics of the specific respiratory event.

### F. Performance of AHI Estimation
#### 1) Comparison of AHI between ROSA and PSG
We can estimate the value of AHI for each subject according to the sleep staging and SAE detection results by ROSA. As introduced in Section III-C, AHI is the ratio of the number of SAEs throughout the night to the TST. TST is calculated from the sleep staging results, and the number of SAEs can be counted according to the SAE detection results. Specifically, we estimate the TST by calculating the duration of Sleep stage, and count the number of SAEs detected during sleep.

The comparison of the estimated and PSG-derived AHI across different methods is shown in Fig. 9. The figure illustrates that methods relying solely on SpO$_2$ or radar signal for SAE detection tend to underestimate the AHI. This is because some types of SAEs, such as CA, do not always coincide with significant OD, and radar signals are prone to missed detections due to motion artifacts or environmental interference. ROSA effectively mitigates these limitations by leveraging the complementary information from both radar and SpO$_2$ signals, thereby improving the detection performance. It demonstrates that ROSA can have a high agreement (ICC=0.9870) with PSG in AHI estimation. We also assess the diagnostic performance of ROSA using AHI thresholds of 5, 15 and 30 events/h, as shown in Table V. It is evident that ROSA exhibits outstanding diagnostic performance for OSAHS, exceeding 88% in sensitivity, specificity, and accuracy across all thresholds. These results demonstrate the effectiveness of ROSA in OSAHS diagnosis and highlight its clinical value.

#### 2) Comparison with AHI Estimation of Previous Studies
Table VII presents the performance of recent studies on radar-based or SpO$_2$-based AHI estimation on their respective datasets. Our study shows better performance than all these studies, as evidenced by the ICC and Pearson correlation coefficient ($r$). Additionally, the larger sample size in our study compared to those radar-based studies increased the confidence of our results.

#### 3) OSAHS Severity Classification
We used the estimated AHI values to classify the severity of

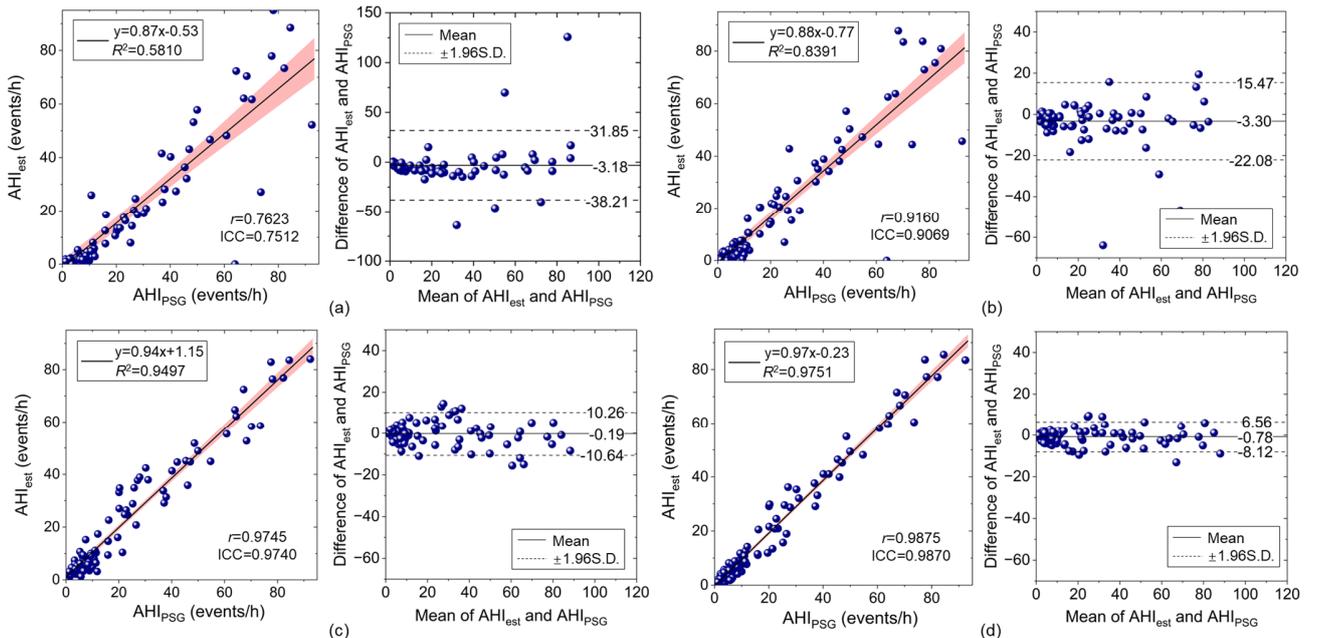

Fig. 9. The scatter plot and Bland-Altman analysis of the estimated and PSG-derived AHI across different methods. $R^2$, $r$ and ICC values are given. Regressions based on the data are also plotted. (a) OxiNet; (b) ODI$_3$; (c) RASA R-CNN; (d) ROSA.

TABLE VII
COMPARISON OF AHI ESTIMATION PERFORMANCE WITH RELATED STUDIES

| Study | Sensor | N | Performance ICC | r |
|---|---|---|---|---|
| Nikkonen et al. [36] | Oximeter | 198 | 0.960 | / |
| Levy et al. [34] | Oximeter | 562 | 0.960 | / |
| Hasan et al. [35] | Oximeter | 478 | / | 0.970 |
| Wei et al. [13] | UWB | 67 | / | 0.820 |
| Choi et al. [16] | FMCW | 35 | 0.889 | 0.892 |
| Kang et al. [14] | IR-UWB | 94 | 0.927 | 0.960 |
| Choi et al. [15] | FMCW | 44 | 0.929 | 0.949 |
| Kwon et al. [17] | IR-UWB | 36 | / | 0.970 |
| RASA R-CNN | FMCW | 100 | 0.974 | 0.975 |
| ROSA | FMCW & Oximeter | 100 | 0.987 | 0.988 |

RF, Radio frequency; FMCW, Frequency-modulated continuous-wave; IR-UWB, Impulse-radio ultra-wideband; N, Number of subjects; ICC, Intraclass correlation coefficient.

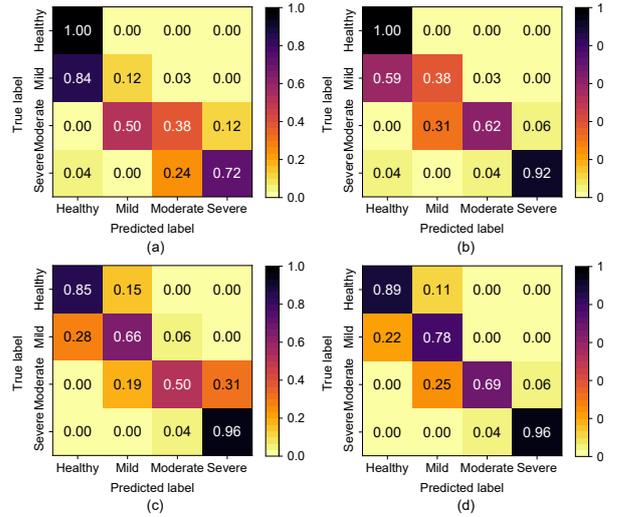

Fig. 10. Confusion matrix for OSAHS severity classification. (a) OxiNet; (b) ODI$_3$; (c) RASA R-CNN; (d) ROSA.

OSAHS in the subjects and compared the results with those obtained from PSG. Fig. 10 presents the confusion matrix for OSAHS severity classification based on different methods. As shown in the figure, ROSA can accurately assess the subjects' condition, with only a few misclassifications occurring between adjacent severity levels. The classification results based solely on SpO$_2$ data correctly identify all healthy subjects. However, they suffer from an overall underestimation of AHI values, leading to poor severity classification performance. The confusion matrix generated from the RASA R-CNN shows inferior performance compared to ROSA in distinguishing between mild and moderate OSAHS subjects.

## IV. DISCUSSION

This paper introduces ROSA that diagnoses OSAHS using only millimeter-wave radar and pulse oximeter. It estimates TST through a sleep staging network and detects SAE through an RCNN-based network. It also utilizes the decision-level fusion strategy to integrate the information from both radar and SpO$_2$ signal. The AHI value is then calculated for the diagnosis of OSAHS. Experimental results demonstrate that ROSA achieves high agreement with PSG in AHI estimation and OSAHS diagnosis. This shows the reliability of ROSA for monitoring sleep-disordered breathing, and its potential to promote accessible sleep monitoring.

In contrast to previous studies, we propose the decision-level fusion strategy to improve the performance of SAE detection. Experimental results have demonstrated the superiority of fusion in improving SAE detection performance. This is largely due to the capacity of neural network to effectively learn complex patterns and relationships within the data. Analysis of the results before and after fusion reveals that the primary role of SpO$_2$ is to reduce false alarms in the SAE detected by RASA R-CNN. However, the fusion of radar and SpO$_2$ signals does not significantly increase the recall rate of SAE due to the drawbacks of the fusion strategy and the inherent inability of SpO$_2$ to independently locate SAEs.

Additionally, in RassNet, we introduce a change loss function and a two-stage training strategy. Results in Table III show that the change loss leverages the continuity of sleep stages, effectively improving the performance of WS classification. The two-stage training strategy reduces the complexity of learning transition relationships for the network, making it easier for the CRF layer to capture the transitions between different stages. However, the CRF loss does not have a positive effect on WS classification, because the transitions between wakefulness and sleep do not follow a predictable pattern.

As illustrated in the confusion matrix shown in Fig. 10, ROSA exhibits a tendency to misclassify subjects categorized as mild and healthy, as well as those classified as mild and moderate. On the one hand, this shows that ROSA still has shortcomings in AHI estimation. On the other hand, it is worth noting that the severity of OSAHS is graded by different AHI thresholds. Therefore, small errors in AHI estimation near these thresholds may lead to incorrect severity classification.

Although ROSA has many advantages, it also has several limitations. First, the proposed fusion strategies primarily adjust the confidence of SAEs detected by RASA R-CNN with blood oxygen information, which can hardly recall abnormal events missed by radar. Secondly, ROSA is less effective in classifying different types of SAEs, as shown in Table VI. PSG identifies the specific type of SAE by analyzing its underlying causes, which necessitates the use of respiratory airflow and respiratory effort signals. However, to facilitate sleep monitoring with low-burden device, ROSA does not incorporate respiratory airflow signals, resulting in reduced capability for distinguishing among different event types.

To enhance the effectiveness of ROSA, several future directions for research and development should be considered. Future research should focus on further developing the fusion strategy. Specifically, it is essential to thoroughly investigate the information contained in SpO$_2$ signals besides features related to OD events. It is also necessary to utilize SpO$_2$ signals to recall SAE missed by RASA R-CNN. Without significantly increasing the burden of subjects, employing new sensors to collect other physiological signals can benefit comprehensive sleep assessments. Additionally, conducting experiments with larger sample sizes is necessary to enhance the robustness and reliability of ROSA.

## V. Conclusion

In this study, we introduce a low-burden method, ROSA, designed for the diagnosis of OSAHS utilizing millimeter-wave radar and pulse oximeter. This method performs sleep staging and directly localizes SAEs temporally without the need for post-processing, enabling accurate estimation of AHI without additional support. ROSA effectively integrates information from both radar and $SpO_2$ signals to enhance SAE detection accuracy. Experimental results on a real-world dataset indicate a high agreement between ROSA and PSG, demonstrating robust diagnostic capability. Comparative analysis with prior studies further highlights the advantages of ROSA. ROSA offers an efficient solution for OSAHS diagnosis using low-burden devices, showing significant potential to improve diagnostic accessibility for OSAHS.

## Acknowledgment

This work was supported in part by the National Natural Science Foundation of China under Grant 61925106